\documentclass{SICFP23}

\usepackage{subfigure}
\usepackage{latexsym}
\usepackage{xspace}
\usepackage{authblk}

\DeclareMathAlphabet{\bm}{T1}{ptm}{b}{it}

\usepackage{amsmath,amssymb,amsfonts}
\usepackage{bm}
\usepackage{graphicx}
\usepackage{textcomp}
\usepackage{xcolor}
\usepackage{enumerate}

\usepackage{booktabs,siunitx,multirow,enumitem} 
\usepackage{subfigure}
\usepackage{hyperref}
\usepackage{cite}
\usepackage{color,soul}
\newcommand{\DONE}[1]{#1} 

\title{Towards Energy Efficient Control for Commercial Heavy-Duty Mobile Cranes: Modeling Hydraulic Pressures using Machine Learning}
\author[1,2]{Abdolreza Taheri}
\author[1]{Robert Pettersson}
\author[1]{Pelle Gustafsson}
\author[3]{Joni Pajarinen}
\author[2]{Reza Ghabcheloo}
\affil[1 ]{Control Systems R\&D, HIAB, Hudiksvall, Sweden}
\affil[ ]{\tt \{abdolreza.taheri, robert.pettersson, pelle.gustafsson\}@hiab.com}
\affil[2]{Faculty of Engineering and Natural Sciences, Tampere University, Tampere, Finland}
\affil[ ]{\tt \{reza.taheri, reza.ghabcheloo\}@tuni.fi}
\affil[3]{Department of Electrical Engineering and Automation,
        Aalto University, Espoo, Finland}
\affil[ ]{\tt joni.pajarinen@aalto.fi}


\begin{document}


\maketitle
\cfoot{\thepage}
\begin{center}
  \begin{minipage}{12cm}
    \begin{abstract}
      A sizable part of the fleet of heavy-duty machinery in the construction equipment industry uses the conventional valve-controlled load-sensing hydraulics. Rigorous climate actions towards reducing CO$_{2}$ emissions has sparked the development of solutions to lower the energy consumption and increase the productivity of the machines. One promising solution to having a better balance between energy and performance is to build accurate models (digital twins) of the real systems using data together with recent advances in machine learning/model-based optimization to improve the control systems. With a particular focus on real-world machines with multiple flow-controlled actuators and shared variable-displacement pumps, this paper presents a generalized machine learning approach to modeling the working pressure of the actuators and the overall pump pressures. The procedures for deriving reaction forces and flow rates as important input variables to the surrogate models are described in detail. Using data from a real loader crane testbed, we demonstrate training and validation of individual models, and showcase the accuracy of pressure predictions in five different experiments under various utilizations and pressure levels.
    \end{abstract}
  \end{minipage}
\end{center}


\section{Introduction}\label{sec:Introduction}
Heavy-duty machines are the key assets for productivity in today's industry. A large portion of heavy-duty machines are driven by multiple hydraulic actuators on linked joints. In the conventional load-sensing (LS) valve-controlled configurations that are prevalent in excavators and loader cranes, the actuators operate on different pressure levels and are often supplied by one or more shared variable-displacement pumps. In such configuration, moving two or more joints on the machine would raise the pressure level of the pump to match the demand of the actuator with the highest operating pressure. This will result in losses in every other actuator that is simultaneously being utilized but has a signal pressure lower than the maximum pressure. Therefore, one trivial recommendation to avoid these so-called throttling losses has been to utilize actuators sequentially during control tasks, especially in cases where the pressure levels tend to vary significantly between actuators. However, such strategy would greatly restrict the motion of the machine and therefore results in a higher total time-to-reach, since only one joint is moving at any given time instance. Given that the total energy expenditure of the combined functions in a machine not only depends on the actuator pressure levels, but how they are utilized (the control strategy), it has remained an open question how to develop a reliable control system that minimizes not just the throttling losses, but the total energy for a maneuver during multi-joint motions of a real machine.

Generally, improving the energy efficiency of conventional heavy-duty machines is being tackled in two directions. On one front, there are studies proposing alternative concepts that are believed to be more efficient which require replacing lots of hardware components or completely rebuilding the system \cite{qu2021high,fassbender2020finding}. In many industrial applications, the high incurred cost of these novel concepts or their difference in performance compared to conventional hydraulics has made it too difficult for the industry to switch the hardware \cite{fassbender2021improving}. Consequently, research and development has focused on the second direction of improving the energy efficiency: optimizing the components or parameters in the conventional systems~\cite{fassbender2021improving, bedotti2018modelling} and developing intelligent energy-optimized controllers~\cite{nurmi2017global, ZHENG2021113762} which cost less and could potentially improve the current fleet of heavy-machines as well.
A modeling and investigation of various system layouts for a hydraulic excavator is presented in \cite{bedotti2018modelling}, which also includes the load-sensing scheme studied in this paper. It has also been shown, that not only the energy consumption can vary between different layouts, but there is also the possibility to optimize flow areas using genetic algorithm to minimize fuel consumption in \DONE{working} cycles.

The usual approach to achieving energy balance in optimal control methods and reinforcement learning is to include the norm of control signals $|u|$ or joint velocities in the cost function~\cite{siliciano2010robotics}. Unfortunately, this approach does not work for systems where actuator losses are interdependent, e.g. in LS hydraulics with a shared variable-displacement pump, where the working pressure of one actuator can raise the system-level pressure and result in high throttling losses in all active cylinders in the system. Very few works have investigated efficient control of heavy-duty machines using the hydraulic energy consumption as an objective. Among them, a solution to the constant-pressure redundant hydraulic manipulator in simulation has been developed in~\cite{nurmi2017global} using the dynamic programming algorithm. \DONE{Our work is motivated by the fact that the total hydraulic energy consumption is a better metric for energy optimization than minimizing just the throttling losses, and is an objective that can be modeled and predicted by data-driven techniques. This is certainly a step in developing predictors of energy in heavy-duty hydraulic machines, which can be further augmented with models of other energy-consuming parts in the machine, such as the pump and motor efficiencies. In particular, we propose a method for training and testing on data gathered directly from a real loader crane with redundant configuration,} in which the (variable-displacement) pump pressure levels are constantly varying based on actuators' utilization. Similarly, a dynamic programming approach to optimal energy motion of redundant hydraulic manipulators for constant-pressure and LS systems has been developed in~\cite{ZHENG2021113762}. We consider systems in which the pressure model is not perfectly known, and aim to learn these models from real machine data. The proposed models are compatible for use in gradient-based controller learning algorithms~\cite{taheri2022BAGEL}, they are faster in reaching a solution compared to dynamic programming, do not require quantization of states and actions, and scale better with respect to the number of dimensions.

Data-driven and machine learning methods have shown great success in modeling complex, non-linear dynamical systems. The resulting models have many practical use cases in engineering systems: They can be used to validate the behavior of system components, to train and deploy inverse models and feedforward controllers by training on real machine data as recently shown for spool valve hydraulics in~\cite{taheri22RAL}. Surrogate models (also referred to as "digital twins") can also be used for health monitoring of real systems, to warn against sudden discrepancies in a system's behavior. In~\cite{TANG2022106300}, a method of fault analysis for hydraulic pumps is proposed based on an adaptive convolutional neural network (CNN) deep learning architecture. Pressure prediction for a single-actuator variable-speed pump controlled testbed is studied in~\cite{kilic2014pressure} using a structured recurrent neural network (RNN) model. Moreover, in many cases the models are differentiable and can be used for optimization of intelligent control systems using gradient-based approaches~\cite{taheri2022BAGEL}. However, the technology is still maturing to achieve the ultimate goal of having real-world applicable intelligent control systems that can predict and minimize the complex energy consumption of the multi-actuator machines throughout operations. In order to achieve this goal, it is required to have accurate and reliable predictive models for the pressure levels in the actuators and pumps.

In our work, we aim to address this gap by proposing data-driven predictive models of actuators' working pressure and the pump pressure in loader cranes. The methodology is discussed in detail in sec.~\ref{sec:method}, along with a thorough overview of the loader crane kinematic transformations and load dynamics that are used for calculation of important input features. The models are validated on a real 21 T.m. loader crane system with three links and a variable-displacement pump, the results of which are presented and discussed in sec.~\ref{sec:results}. Finally, a conclusion is drawn based on the results of this study in sec.~\ref{sec:conclusion}.

\section{Method} \label{sec:method}
\begin{figure}[!tp]
  \centering
  {\includegraphics[width=0.99\columnwidth]{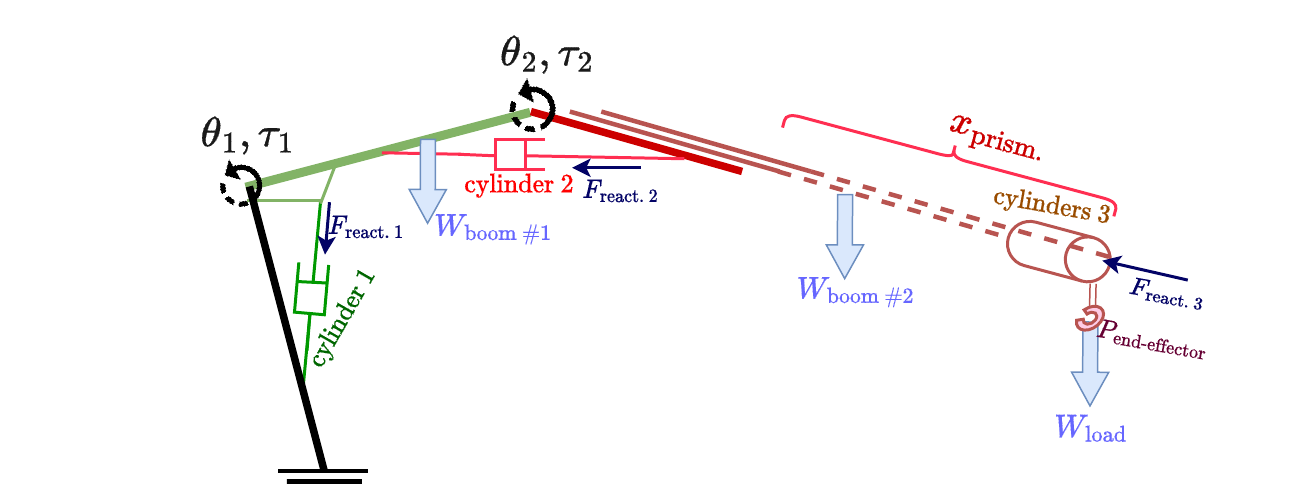}}
  \caption{Schematic of the heavy-duty loader crane testbed. The system comprises two revolute joints ($\theta_1, \theta_2$) and one prismatic joint ($x_\textrm{prism.}$) actuated by six serially connected double-acting hydraulic cylinders. The position of end-effector and links' center of gravity can be calculated using kinematics. The weight of components and the end-effector load create reaction torques at the joints and consequently reaction forces on the cylinders, which affects the working pressure of hydraulics.} 
  \label{fig:sketch}
\end{figure}

This section details our approach to modeling and estimation of the pump pressure in a load-sensing pressure-compensating (LSPC) system with a variable-displacement pump. Specifically, we describe the calculation of dynamic variables for a loader crane model that is actuated by two cylinders on revolute joints and multiple cylinders acting on a long prismatic link, as shown in fig.~\ref{fig:sketch}. The actuators share a variable-displacement pump and are commanded through a directional control valve (DCV). It is important to point out that there are no model-specific assumptions in our methodology that could prevent the approach from scaling to machines with different kinematics and different number of actuators or shared pumps.

In the sections that follow, we start off by deriving the building blocks of the dynamic features that are used as input to the surrogate pressure models. The system dynamics and variables are overviewed in sec.~\ref{sec:sys_dyn}, focusing on the crane testbed shown in fig.~\ref{fig:sketch}. Section~\ref{sec:force} details the calculation of reaction forces on individual cylinders and sec.~\ref{sec:flow} overviews the pressure and flow dynamics in the actuators. Section~\ref{sec:pres13} describes the use of these dynamic variables for optimizing the machine learning models so as to predict the working pressure of individual actuators, which are the deciding factors in overall pump pressure of the system that is modeled in sec.~\ref{sec:p_pump}.

\subsection{Overview of System Dynamics and Variables} \label{sec:sys_dyn}

The total expended energy in a machine's hydraulic functions is proportional to the total flow and pump pressure throughout a motion. These dynamic properties vary depending on many factors, most significantly on the utilization of actuators and the crane state. The utilization of actuators is the command signal $u_\textrm{cmd}$ (current input) to spool valves~\cite{taheri22RAL}, the state of the actuators is defined by the displacement of spool $x_{s}$ and the position of piston $x_p$ and side pressures $\{P_A, P_B\}$ in each hydraulic cylinder. The state of crane is the angle/position of revolute/prismatic joints $\{\theta_1, \theta_2, x_\textrm{prism.}\}$ (known collectively as the crane pose) and their rate of change $\{\dot{\theta}_1, \dot{\theta}_2, \dot{x}_\textrm{prism.}\}$, in addition to variables that affect the joint torques such as load weight $W_\textrm{load}$ on the end-effector.

The kinematics of the crane, as typically described by geometric parameters of links within the Denavit-Hartenberg (DH-) convention\cite{siliciano2010robotics}, relates any position on crane links (e.g. position of the end-effector $x_{ef}$, center of gravity of link weights $x_{cg_i}$ and load shown in fig.~\ref{fig:sketch}) to the joint states by a general transformation and rotation. The rate of change of these variables is associated by the Jacobian of the transformations $\bm{J}{(\Theta)}$ with respect to joint states: 
\begin{equation}
  \label{eq_Jacobian}
  \bm{v}_{ef} = \frac{d\bm{x}_{ef}}{dt} = (\frac{\partial \bm{x}_{ef}}{\partial{{\Theta}}}) \frac{d{{\Theta}}}{d{t}} = \bm{J}_{_{ef}}{(\Theta)} \dot{{\Theta}}
\end{equation}
where $\Theta = [\theta_1, \theta_2, x_\textrm{prism.}]$ is the vector of joint variables.

The transformation from the crane joint states to the cylinders and vice versa is done by a (non-linear) mapping function $\mathcal{C}(.)$ which is calculated from the geometry of the cylinder placement. With this mapping, the piston positions and joint states relate to each other via:
\begin{align}
  \label{eq_cyl_len}
  x_{p_i}  & = \mathcal{C}_i(\theta_i)                           \\
  \theta_i & = \mathcal{C}^{-1}_i(x_{p_i})\label{eq_cyl_len_inv}
\end{align}
The relationship between the rate of change of these states can be obtained via differentiation and chain rule:
\begin{align}
  \label{eq_cyl_spd}
  \frac{dx_{p_i}}{dt} & = \frac{\partial {\mathcal{C}_i}}{\partial{\theta_i}}\Bigr|_{\substack{\theta_i={\theta}_{i}(t_k)}} \frac{d\theta_i}{dt} 
\end{align}
Note that the mapping $\frac{\partial {\mathcal{C}_i}}{\partial{\theta_i}}$ assumes strictly positive values in crane's operational envelope, and hence the inverse mapping from the actuator speeds to joint speeds exists (the same is true for mapping the joint torques to cylinder reaction forces via eq.~\eqref{eq_Freact}, detailed in section~\ref{sec:force}).

Within this framework, all state variables and their rate of change are defined by deterministic functions that are backwards-differentiable via the chain rule. Having differentiable functions enables controller optimization using recent model-based algorithmic advances~\cite{taheri2022BAGEL}. In a similar fashion, the modeling of the flow, forces, and pressures described in the succeeding sections will incorporate backwards-differentiable models to make it possible to compute the gradients of the total energy expenditure objective for energy-efficient controller optimization.

\subsection{Calculation of Cylinder Forces}\label{sec:force}

The weight of each link and the weight of the load on the end-effector will cause reaction torques $\{\tau_1, \tau_2\}$ on the loader crane joints as shown in fig.~\ref{fig:sketch}. The amount of joint torques with respect to the generalized force contributions $\bm{\gamma}_{w_i}$ of a weight component can be calculated by the principle of virtual work~\cite{siliciano2010robotics}:
\begin{align}
  \label{eq_Vwork}
   & \delta W_\tau = \bm{\tau}^T \delta \Theta                        \\
   & \delta W_\gamma = \bm{\gamma}^T   \bm{J}{(\Theta)} \delta \Theta \\
\end{align}
which, at static equilibrium ($\delta W_\tau = \delta W_\gamma$) becomes:
\begin{align}
   & \bm{\tau} = \bm{J}^T(\Theta)\bm{\gamma}
  \label{eq:torq}
\end{align}
and the torque on each joint is calculated by accumulating individual contributions of weight forces on that joint. Equation~\eqref{eq:torq} shows that the transpose of the Jacobian defines the transformation from weight forces to joint torques. More importantly, the total reaction torque (and consequently, the reaction forces on cylinders) can be viewed as a single feature that embeds the Jacobian transformation as well, which is evident in eq.~\eqref{eq:torq}. For this reason, the pressure models will be able to make predictions based on reaction forces without explicit access to crane states or the Jacobian.

Finally, the force reaction on each cylinder due to the (static) load and boom weights can be obtained by the inverse of the partial differences relationship between the joint variables and cylinder states, derived in eq.~\eqref{eq_cyl_spd}:
\begin{equation}
  \label{eq_Freact}
  F_\textrm{s. react.,i} = ( \frac{\partial \bm{x}_{p_i}}{\partial \theta_i} )^{-1} \tau_i
\end{equation}

In addition to the calculated static load forces, there are dynamic forces (such as friction, load inertias, etc.) that come into effect during the utilization of actuators. Owing to high forces across the actuator and fast pressure transients, the actuator is considered as a quasi-steady state process (see~\cite[ch. 6.3]{manring2019hydraulic}), i.e. retracting and extending occurs at constant speeds with near-zero net force. As a result, the combined effect of the static and dynamic forces, or the total reaction force, can be calculated from the pressure difference across the two sides of the piston during motion, using the following relation:
\begin{equation}
  \label{eq_Freact_total}
  F_\textrm{total, i} = P_A \mathcal{A}_A - P_B \mathcal{A}_B
\end{equation}
where $P$ and $\mathcal{A}$ denote the pressure and area of each side of the actuator.

\subsection{Cylinder Pressure and Flow Dynamics} \label{sec:flow}

In conventional approaches to modeling hydraulics, the side pressures are typically formulated as ordinary differential equations that describe their rate of change (see~\cite[ch. 4.2.2 \& eqs.~(94)-(95)]{heinze2008modelling}), e.g.:
\begin{equation}
  \label{eq_dp}
  \frac{dP_{A/B}}{dt} = \frac{E_\textrm{oil}}{V_{A/B}}\left( Q_{A/B} + Q_\textrm{in/ex} - \mathcal{A}_{A/B} \frac{dx_p}{dt}\right)
\end{equation}
where $Q_\textrm{A/B}$ is the flow, $Q_\textrm{in/ex}$ denotes the internal/external leakages, $V_{A/B}$ denotes the volume in chambers $A$ or $B$, and ${E_\textrm{oil}}$ is the bulk modulus for the oil.

The rate of change in the state of an actuator in an LSPC system considering quasi-static conditions with negligible pressure transients ($\frac{dP}{dt} \approx 0$ in eq.~\eqref{eq_dp}), according to a previous detailed study~\cite{taheri22RAL} is proportional to the flow rate:

\begin{equation}
  \label{eq_cyl_rate}
  \dot{x}_{p_i} = f_\textrm{cyl. rate}(Q_i, \bm{\varphi}_i)
\end{equation}

where $Q_i$ denotes the active flow from the spoolvalve to the $i^{th}$ cylinder and $\bm{\varphi}_i$ is cylinder-specific structural parameters, i.e. the piston and rod diameters. On revolute joints, the cylinder states and rate of change of states are measured with sensors on the joint angles, then converted according to eqs.~\eqref{eq_cyl_len}-\eqref{eq_cyl_spd}. It is also standard practice to use a filter (e.g. Savitzky-Golay \cite{savitzky1964smoothing}) to obtain derivatives or just to improve the quality of the measurements~\cite{taheri2022BAGEL}.

We consider cylinders with different side areas $\{\mathcal{A}_{A}, \mathcal{A}_{B}\}$, so given the same actuator speed $\dot{x}_p$ there will be different flow rates on the sides $\{{Q}_{A}, {Q}_{B}\}$, which can be calculated using the inverse of eq.~\eqref{eq_cyl_rate}, i.e.

\begin{equation}
  \label{eq_cyl_flow}
  Q_i =  f^{-1}_{\bm{\varphi}_i,_\textrm{cyl. rate}}(\dot{x}_{p_i})
\end{equation}

For the purpose of modeling side pressures later in sec.~\ref{sec:pres13} we will treat the flow rates and parameters for each side separately, i.e. $\{\bm{\varphi}^+_i, Q^+_i\}$ during cylinder extension and $\{\bm{\varphi}^-_i, Q^-_i\}$ during retraction.

Since eq.~\eqref{eq_dp} is a rather simplified model with heuristically estimated parameters, we do not explicitly take it as the model for pressures. However, it is important to point out the variables that affect the pressure for the purpose of designing a machine learning model. Flow rate is evidently one of the main decision variables since flow-dependent losses are an important factor in the pressure dynamics.
The cylinder speed $\dot{x}_p$ according to section~\ref{sec:flow} has a direct dependency on the flow $Q_\textrm{A/B}$. Additionally, the chamber volume can be calculated from the side areas and position of the piston, which is related to crane states by the kinematic transformation in eq.~\eqref{eq_cyl_spd}. We can also assume, that the leakage flow \DONE{(which generally has smaller contribution to pressure drops \cite{heinze2008modelling}) }is a non-linear function of the cylinder speed $\dot{x}_p$ and the pressure difference across sides, which makes it proportional to both flow rate and reaction force. Moreover, the friction forces that make up eq.~\eqref{eq_Freact_total} can be modeled as a function of the load reaction forces (eq.~\eqref{eq_Freact}) derived from the crane pose, as well as a function of the cylinder speed (i.e., flow rate)~\cite[ch. 5]{heinze2008modelling}.

\subsection{Actuator Working Pressure Models}\label{sec:pres13}

The analysis described in sec.~\ref{sec:flow} identifies two important variables (flow and forces) that we will use as input signals to the working pressure models. The output of the models are the actuators' pressure demand that is signaled to the pump to be supplied (as will be utilized in sec.~\ref{sec:p_pump}). For each actuator, the model consists of two Gaussian Processes (GPs) that predict the absolute pressure value for either side of the piston (denoted by $p^{\ast +}_i$ and $p^{\ast -}_i$). The input features $\{F_\textrm{s. react.,i},Q_i\}$ to each of these models are calculated using the equations for static force~\eqref{eq_Freact} and flow rate~\eqref{eq_cyl_flow} that were discussed earlier in sections~\ref{sec:force} and~\ref{sec:flow}, respectively. GPs are probabilistic machine learning models that are widely used in real-world robotics applications since they require small training data compared to other machine learning models such as neural networks (for more details, see~\cite{taheri22RAL,taheri2022BAGEL}).
The reason for having two separate inner models is inspired by the mechanism of pressure signals from the actuators and primary shuttle valves to the pump\cite{dell2017}. That is, depending on the direction of the flow of a single actuator, one of the two side pressures decides the working pressure $P_i$ that should be supplied by the pump (also referred to as signal pressure~\cite[ch.~18]{dell2017}). Having two GPs to describe the two side pressures makes it easier to train the models since data for each side is available by separate sensor measurements. After defining the models, the training dataset of inputs ($\bm{Q}_i, \bm{F}_\textrm{s. react.,i}$) and outputs $\bm{P}_i$ for each actuator on the real system are selected according to the direction of motion of cylinder (superscripted with~$^+$ and~$^-$, refer to sec.~\ref{sec:flow}) and labeled as working pressure. When the DCV is in neutral position and the cylinder has no motion, the working pressure defaults to zero since no pump supply is needed. Therefore, the model for predicting the $i$th actuator's working pressure at new test inputs $\{Q^\ast_i, F^\ast_\textrm{s. react.,i}\}$ takes the following form:
\begin{equation} 
  \label{eqn_p13}
  P_i(Q^\ast_i, F^\ast_\textrm{s. react.,i}) =  \begin{cases}
    p^{\ast +}_i | (\bm{Q}^+_i, \bm{F}^+_\textrm{s. react.,i}, \bm{P}^+_i,Q^\ast_i, {F}^\ast_\textrm{s. react.,i}) \sim {GP}(\mu^{\ast +}_i,  \Sigma^{\ast +}_i  ) ,       & \dot{x}_{p_i} > 0 \\ 
    0,                                                                                                                                                                     & \dot{x}_{p_i} = 0 \\
    p^{\ast -}_i | (\bm{Q}^{-}_i, \bm{F}^{-}_\textrm{s. react.,i}, \bm{P}^{-}_i,Q^\ast_i, {F}^\ast_\textrm{s. react.,i}) \sim {GP}(\mu^{\ast -}_i,  \Sigma^{\ast -}_i  ) , & \dot{x}_{p_i} < 0
  \end{cases}
\end{equation}
where ($\mu^{\ast}_i,  \Sigma^{\ast}_i $) denote the GP posterior mean and covariance evaluated at the test inputs. For a more detailed formulation of these variables, refer to~\cite[ch. 2.2]{williams2006gaussian} or the previous work by authors~\cite[sec. III-A]{taheri2022BAGEL}.


\subsection{Pump Pressure Model}\label{sec:p_pump}
In many commercial flow-controlled heavy-duty machines, the hydraulic circuits are designed to have one or more shared pumps supply multiple actuators simultaneously. Each actuator entails a minimum required pressure for operating that must be supplied by the pump as soon as the flow opens up to the actuator. The overall pressure requirement depends on factors such as the type and size of the actuator, and as studied in sec.~\ref{sec:pres13} to the target speed and the dynamic loads on the cylinder.

In load-sensing (LS) systems in particular, as the working pressure levels for each actuator vary during a trajectory, the (variable-displacement) supply pump pressure is adjusted accordingly to match the highest pressure demand among all the actuators. Understanding the mechanics of the load-sensing system is important from the point of view of control and the machine's energy consumption: if an actuator with a high pressure requirement is commanded with even the slightest flow, the pump pressure level has to rise to match a higher pressure, which in turn causes excessive "throttling losses" in other actuators with lower working pressures that are connected to the same pump.

There are additional considerations concerning the pump pressure model in a real heavy-duty machine. The pump pressure not only has to match the highest of the actuators' pressure demand, but supplies additional pressure above this value as a safety margin to \DONE{make it possible for oil to flow through the system as well as accounting for uncertainties and to }provide better controllability and responsiveness. Since the described pressure margin is an important but unknown parameter of the pump model, we propose to treat it as a parameter that is learned from the real pressure data. Specifically, the output of each of the individual actuator pressure models in sec.~\ref{sec:pres13} are first elevated by the corresponding margin pressure variables $\{c_{P_1}, c_{P_2},..., c_{P_i}\}$, and the overall demanded pressure is set equal to the maximum value of the pressure set:
\begin{align}
  \label{eq_p_demand}
  P_{demand} & = \max\Bigl\{ \left[P_1 + c_{P_1}\mathcal{F}(Q_1)\right], \left[P_2 + c_{P_2}\mathcal{F}(Q_2)\right], ..., \left[P_i+ c_{P_i}\mathcal{F}(Q_i)\right]\Bigr\} \\
             & c_{P_1}, c_{P_2},..., c_{P_i} \geq 0 \nonumber
\end{align}
up to $i$ actuators that are supplied by the same pump. The output of eq.~\eqref{eq_p_demand} provides the pump pressure in the working mode, which \DONE{includes the effects of the secondary shuttle valves and the margin pressure combined\cite[ch. 18]{dell2017}. Note, however, that while this formulation resembles the natural variables of the mechanical system, such as margin pressures, the final values obtained after the optimization process may not reflect the exact values of the real system implementation; rather, they indicate model parameters which result in lowest prediction errors according to the training data and the loss objective. In eq.~\eqref{eq_p_demand}}, $P_i$ is inferred from the data and the probabilistic models in eq.~\eqref{eqn_p13}, and $\mathcal{F}(.)$ is defined as an activation function that determines whether the actuator is in the working mode
\begin{equation}
  \label{eqn_activation}
  \mathcal{F}(Q_i) =  \begin{cases}
    1, & Q_i \neq 0 \\
    0, & Q_i = 0    \\
  \end{cases}
\end{equation}
This ensures that the pressure demands from inactive cylinders remain zero in eq.~\eqref{eq_p_demand}. Unlike the individual actuator pressure models in sec.~\ref{sec:pres13} where the output defaults to zero when the DCVs are in neutral position, the pump maintains a minimum low-pressure when the DCVs are not actuated, called the standby pressure ($P_{standby}$) which is adjusted in the flow control spool\cite[ch. 18]{dell2017}.
Therefore, the final model for the pump pressure becomes:
\begin{align}
  \label{eq_pump_p}
  P_{pump} & = \max\Bigl\{P_{standby}, P_{demand}\Bigr\}
\end{align}

\begin{figure}[!t]
  \centering
  {\includegraphics[width=0.95\columnwidth]{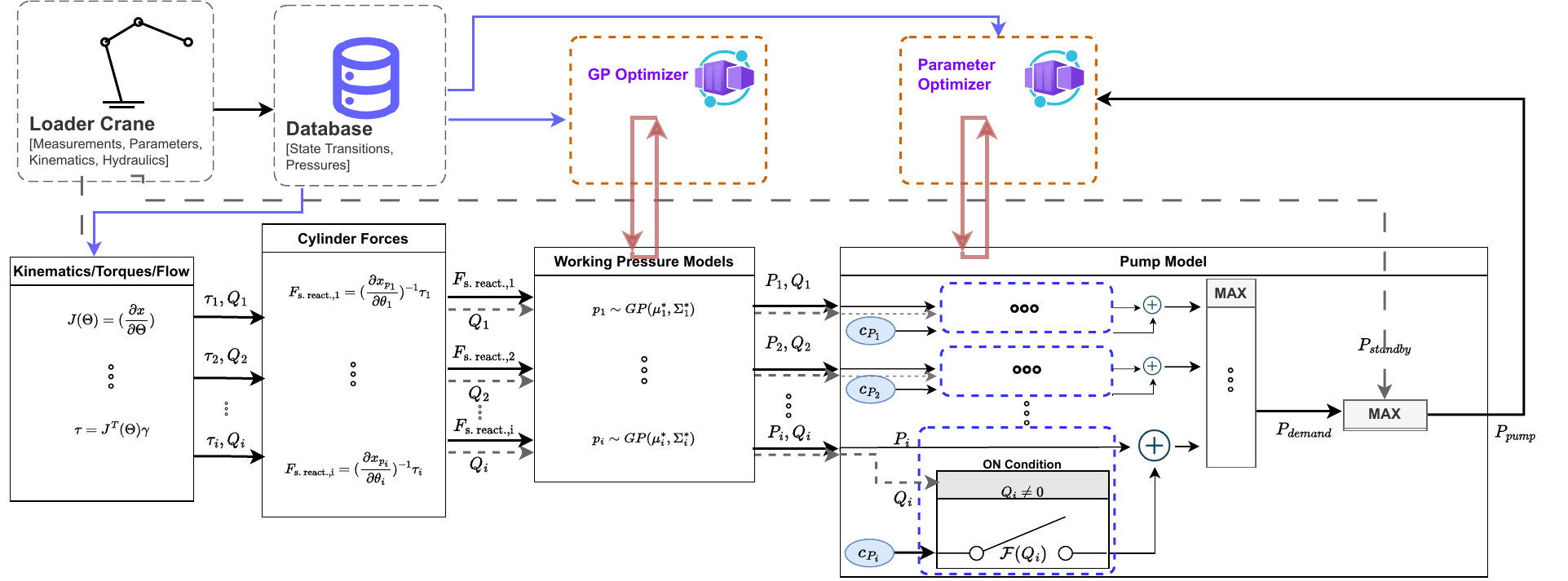}}
  \caption{\DONE{A lean overview of models, data, and optimizers for developing machine learning predictors of working pressure and pump pressure. For more details, refer to the text.}}
  \label{fig:flowchart}
\end{figure}

\DONE{Figure~\ref{fig:flowchart} illustrates the interconnection of the data, calculations, models, and optimization in our proposed training workflow. Now that the pump pressure model is complete}, it is important to elaborate on two main points in the design of the system pressure models: a) The stall mode pressure (when cylinder is at the end of its reach) is not considered in the pump model. This is because the actuator limits are easily handled by the high- and low-level controllers, which can interrupt the command inputs to prevent the pump from reaching stall pressure. As a result, we do not find it useful to include the piston displacement as an extra input to the models to account for effects that do not occur throughout the maneuvers. b) The main reason for training separate models for the actuator working pressures and the pump pressure, is because the $max$ operations in the latter model prevent gradient backpropagation to actuator models that do not have dominating pressure in the system. This results in slow training and will require more data to train the models. On the other hand, since the ground-truth values for each actuator's pressure is available through sensor measurements, it is instead quite seamless to do the training of the models (eqs.~\eqref{eqn_p13}-\eqref{eq_pump_p}) in a decoupled fashion.

\section{Results \& Discussion}\label{sec:results}
\begin{figure}[!t]
  \centering
  {\includegraphics[width=0.98\columnwidth]{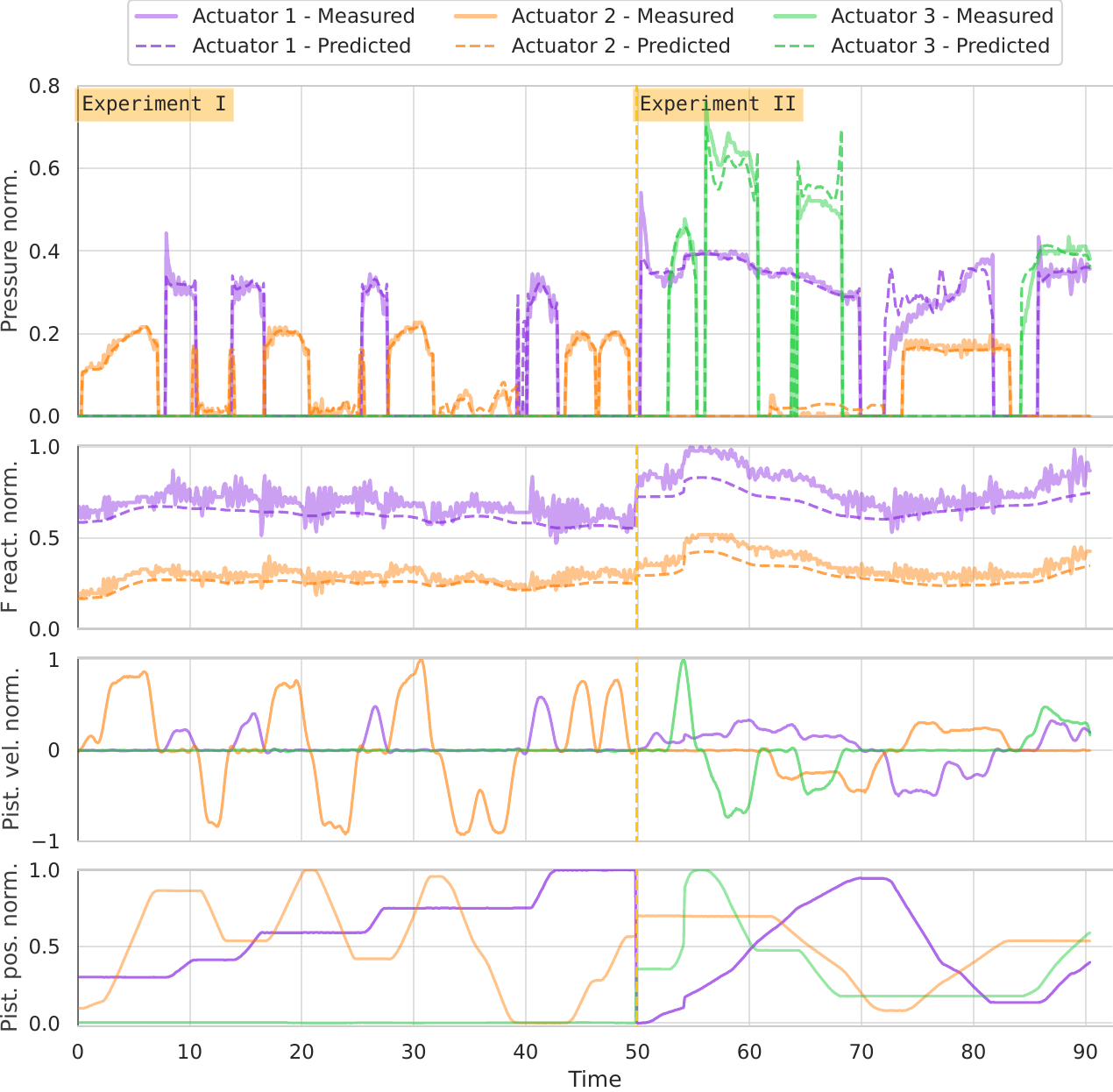}}
  \caption{Experimental results for validation of the three individual actuator pressure models.  From the top, each row indicates: A) Working pressure for each hydraulic actuator (measured vs. model predictions); working/active pressure refers to rod or piston side pressure depending on the cylinder's direction of movement. B) Measured dynamic forces (eq.~\eqref{eq_Freact_total}) vs. calculated static forces from crane pose (as inputs to the pressure models, eq.~\eqref{eq_Freact}) that follow closely the same trend. C) Piston velocity for each actuator, which indicates the cylinders' direction of movement  D) Piston position for each actuator, which alter the crane pose and therefore affect the forces on the cylinders. All values are normalized.} 
  \label{fig:pressure_models}
\end{figure}

\begin{figure}[!t]
  \centering
  {\includegraphics[width=0.98\columnwidth]{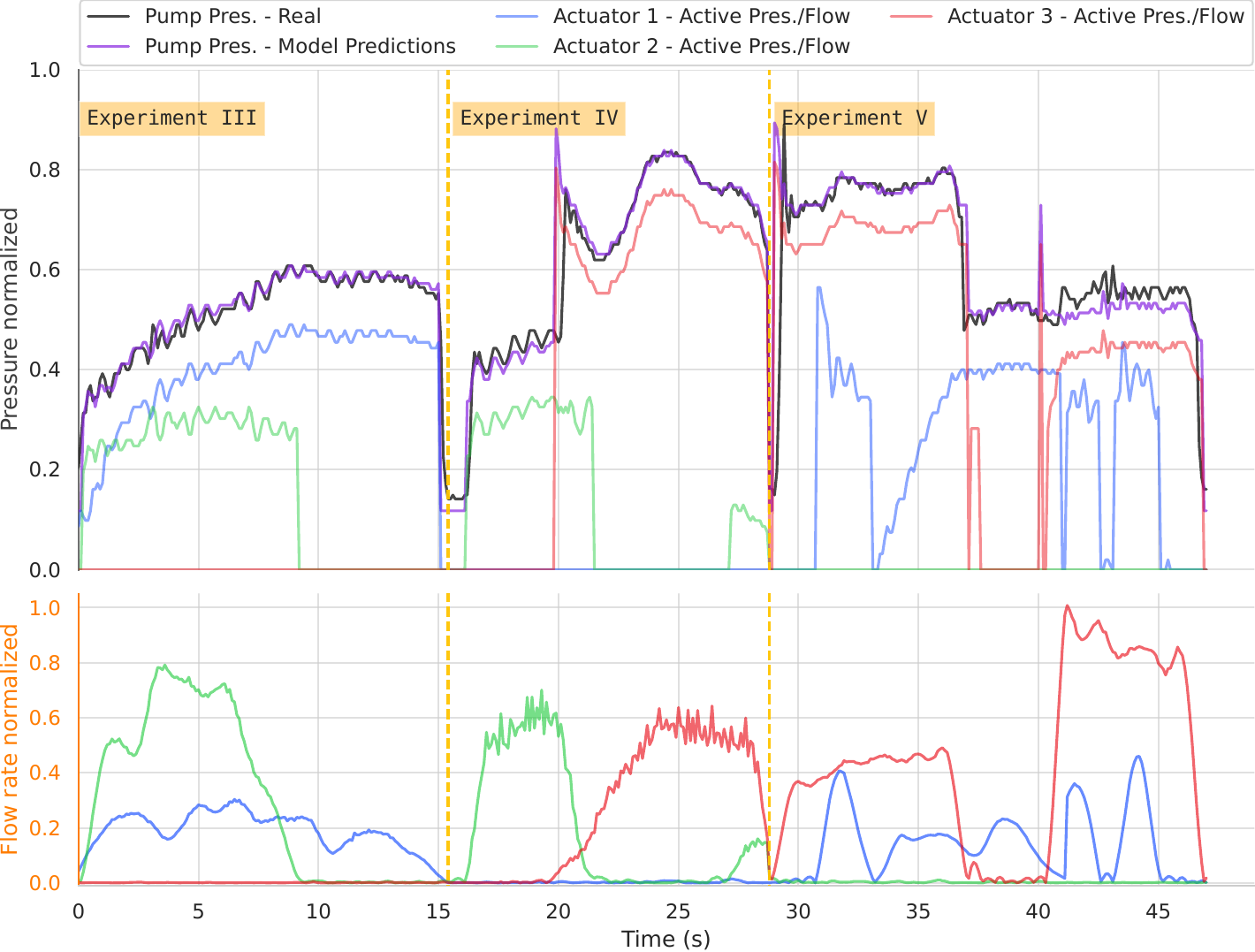}}
  \caption{Experimental results for validation of the pump pressure model. Top: the model predicts the pressure of the pump for the overall system based on the three working pressures as inputs (one from each actuator). Bottom: the history of flow utilization of the actuators that (depending on the crane configuration and reaction forces) result in a different working pressure for each actuator. The actuator with the highest pressure influences the overall pump pressure for the crane.} 
  \label{fig:pump_model}
\end{figure}

This section presents the results of our experiments on a real loader crane to validate the pressure modeling method proposed in sec.~\ref{sec:method}. A total of five tests have been conducted on the loader crane testbed and summarized in figs.~\ref{fig:pressure_models}-\ref{fig:pump_model}. The first two experiments showcase the accuracy of the models for working pressure of individual actuators (as modeled in sec.~\ref{sec:pres13}) as well as the correlation between the calculated and real reaction forces on the actuators (sec.~\ref{sec:force}). The rest of the experimental tests validate the pump pressure model (sec.~\ref{sec:p_pump}) using data from the real loader crane testbed. Since the testbed has three actuators with different sizes and pressure levels, each experiment tries to vary the utilization of actuators to excite different pressure activations in the system.

Figure~\ref{fig:pressure_models} demonstrates the prediction of working pressure models. Each experiment is separated by a dashed vertical line. The measurements from the crane are joint states and actuator/pump pressures from the sensors. Therefore, the ground-truth values for the actuator pressures show real sensor measurements, and the ground-truth values for the forces are calculated from these pressures using eq.~\eqref{eq_Freact_total}. These are all the measurements needed to train the models; whereas after the training, only the joint states information are needed for model predictions. Experiment I focuses on the crane system's revolute joints while the prismatic cylinder set is fixed in the fully retracted position. Actuators 1-2 are then commanded differently to test the pressures under different boom configurations. In Experiment II, all the actuators are jointly moved, including the prismatic cylinders, and take on a high range of configurations. In some cases, e.g. when actuator 2 is retracting, the reaction forces act in the direction of motion, so the actuator's working pressure is zero at some configurations or a small amount at other configurations. One of our main conclusions is that machine learning models are able to predict these variations quite well, though there is no general rule for predicting these effect. Having two separate pressure models for cylinder side pressures is the main contributor to achieve accurate predictions.
\DONE{The experiments are conducted without any external load attached to the end-effector, so that all joints could be driven up to maximum flow and to comply with safety protocols at the indoor laboratory. In the absence of a load on the end-effector, the reaction forces on the prismatic link (actuator 3) are so small that they are excluded in the results shown in fig.~\ref{fig:pressure_models}. Although not validated in our experiments, varying the weight of external loads would result in higher cylinder forces according to eqs.~\eqref{eq:torq}-\eqref{eq_Freact} and, as long as training data in higher force regions are available, the models are expected to predict the pressures accordingly since the weight of end-effector loads are accounted for, similar to link weights (fig.~\ref{fig:sketch}). The weight of the end-effector load, however, is assumed to be known or estimated e.g. by previously developed load-estimation methodologies \cite{ferlibas2021load}.}

It is noteworthy to point out another key result of our design of working pressure models in sec.~\ref{sec:pres13}: \DONE{since the models are trained on the calculated reaction forces (dashed lines, second row of fig.~\ref{fig:pressure_models}), the input values do not necessarily need to correspond exactly to the real forces (solid lines) for the pressure models to work. The real forces (eq.~\eqref{eq_Freact_total}) during these maneuvers are only illustrated for validating the trends that the calculated static forces follow. The discrepancy between these two forces }is mainly attributed to the following factors:
\begin{enumerate}
  \item[a)] Effect of additional weight components, such as the weight of the hydraulic hoses, oil, cylinders and connections, etc. that were neglected or too complex to include.
  \item[b)] Uncertainty in the modeled boom weights or their center of gravity location; especially for the multi-cylinder prismatic joint where individual cylinder positions in serially connected actuators can vary freely during~motions.
  \item[c)] Also, in the modeling of sec.~\ref{sec:pres13}, the dynamic effects (such as friction forces) are not included in the calculated force input to the working pressure models, but instead the flow input to the models complements the information for these effects.
\end{enumerate}

The next experiments (III-V) in fig.~\ref{fig:pump_model} demonstrate the pump pressure model. Similar to the previous tests, effort was made to actuate the cylinders differently to observe variations in the pump pressure. In experiment III, actuator~2 dominates the pressure in the start, then actuator 1 takes over the system's pressure. Experiment IV also starts with actuator 2 but the pump pressure is then taken over by the prismatic cylinders, which collectively have a higher pressure demand. Experiment V shows pump pressure switching starting with actuator 3, then pressure according to actuator 1, and then back to actuator 3, while showcasing that depending on the use case and the crane configuration, the pressure levels between actuators can be very distinct (at the start) so the throttling losses are high, or the pressures can be close to each other (at the end) with less throttling. Moreover, the reported variables $\{c_{P_1}, c_{P_2}, c_{P_3}\}$ \DONE{relating to the effects such as margin pressures, secondary shuttle valves, etc. (refer to sec.~\ref{sec:p_pump}) after the training had significant variation from each other (up to $\pm{20\%}$ of their average value). This confirms our hypothesis that having separate learnable variables for each pressure component results in a more accurate pump model, although given the data-driven nature of all the models, these distinct variables do not necessarily correspond to any of the parameters of the real machine.}


Altogether, we conclude that the trained models are able to make predictions close to the real values, although the inputs neglect the three dynamic uncertainties stated earlier. The models infer these terms from the calculated forces and flow rates and previously observed ground-truth values (i.e., training set), so long as the inputs follow meaningful and consistent trends that reflect the real values for the machine. Under the proposed modeling approach, it is also seamless to learn models of the dynamic forces and treat the weights of the components as learnable parameters, which we will leave for future studies. Finally, the proposed models can be used alongside previously established machine-learning models of flow rates~\cite{taheri22RAL} to estimate the total energy expenditure in a motion. The energy objective can then be directly utilized by a model-based optimization algorithm~\cite{taheri2022BAGEL} to learn energy-optimal high-performing~controllers.

\section{Conclusion}\label{sec:conclusion}
Many recent advances in hydraulic systems, such as modeling the flow in hydraulics and effective controller optimization algorithms through the use of machine learning point towards near future performance improvements for heavy-duty machines. However, optimizing control systems to balance the total energy consumption of hydraulics has been an unsolved problem due to a lack of reliable and differentiable pressure models for real machines.
Our study identifies this gap and proposes an effective machine learning approach to training predictive models of pressure levels for multiple actuators in a load-sensing pressure-compensating (LSPC) hydraulic system with a variable-displacement supply pump. Our analysis demonstrates that the models follow the pressure variations using static forces and flow as decision variables. Moreover, we demonstrated how a pump pressure model with extra learnable parameters can be tuned to accurately predict the overall pressure of the LSPC system.

There are numerous benefits in using machine learning models for predicting system pressures. These models complement the already established models of the flow in the literature to make it possible to estimate the energy expenditure of hydraulic functions. The energy estimation can be incorporated as an optimization objective into gradient-based controller learning approaches, i.e., to optimize control systems for superior energy balance. Furthermore, the pressure models can also be used to optimize or re-design the parameters in the hydraulic system components to improve performance. As highlighted in our discussions, it is also possible to improve the (pressure/force) models to estimate the dynamic forces and the end-effector loads based on the input features during operations, resulting in more intelligent control of heavy-duty machines.

\section*{Acknowledgement}
This work has been funded by the European Union's Horizon 2020 Marie Sklodowska Curie Research and Innovation Programme MORE-ITN under grant agreement No. 858101. The authors gratefully acknowledge Jon Skagersten, Marcus R{\"o}sth, and Szabolcs Fodor at HIAB R\&D for their insights and assistance regarding the loader crane kinematics and hydraulics.

\bibliography{references}
\bibliographystyle{unsrt}

\end{document}